\documentstyle{article}
\begin{document}
\title{Interaction-Free Preparation}
\author{G. Krenn,  J. Summhammer\\
 {\small Atominstitut der \"Osterreichischen Universit\"aten}  \\
  {\small Sch\"uttelstra\ss e 115}    \\
  {\small A-1020 Vienna, Austria   }            \\
  {\small krenn@ati.ac.at}  \\
  {\small summhammer@ati.ac.at}     \\
{\small and}\\
K. Svozil\\
 {\small Institut f\"ur Theoretische Physik}  \\
  {\small Technische Universit\"at Wien   }     \\
  {\small Wiedner Hauptstra\ss e 8-10/136}    \\
  {\small A-1040 Vienna, Austria   }            \\
  {\small svozil@tph.tuwien.ac.at}}
\maketitle

\begin{abstract}
We demonstrate that the preparation of a very well localized atom beam
is possible without physical interaction. The preparation is based on the
selection of an adequate ensemble of atoms of an originally wide beam
by means of information obtained with a neutron interferometer.
In such a case the uncertainty
relation can no longer be interpreted as a by-product of the interaction
between the system and the preparation apparatus.
\end{abstract}

\noindent
\section*{Introduction}
In 1927 Heisenberg formulated the uncertainty relation, which
expresses the fact that the expectation values of
two non-commuting observables cannot be
determined with arbitrary precision. He demonstrated this by means
of a $\gamma$-ray-microscope, which since then has been
discussed in many textbooks of quantum mechanics. In such a
(gedanken-) microscope
the location of an electron is determined by $\gamma$-photons which are
scattered on the electron. Due to the Compton-effect the momentum of
the electron will be changed when the position measurement
(scattering of the photon) takes place.
Because the resolution of the position-measurement is related to the
wavelength of the photons, the momentum transfer in the scattering process
will increase as the accuracy of the position-measurement
is increased.
Therefore it is not possible to
determine both, position and momentum, with arbitrary precision.

This and many other examples which have been invented to illustrate
the meaning of the uncertainty relation may lead to the assumption
that this relation is always based on
a physical interaction between the measured system (electron) and the
system by which the measurement is performed (photon).
This assumption is reasonable when it is assumed
that no measurement is possible without physical
interaction. [Cf. \cite{Gabor}:
measurement by interaction
is associated with the exchange of at least one quantum of action.]
The term ``physical interaction'' is used here for processes which are
associated with the exchange of at least one quantum of action.

The same considerations may also be applied to
the preparation process. In experiments, properties like the spatial
extension or the energy of a system usually are controlled by methods
which imply physical interaction
with the system. Thus the limits in defining the initial conditions
of a system as expressed by the uncertainty relation may again be
interpreted as a consequence of the physical interaction occurring in the
preparation process.

In this paper we will discuss a preparation method which
involves no physical interaction, thereby strictly excluding
such a mechanistic interpretation of the uncertainty relation.
In the proposed setup we use the idea of interaction-free
measurement which has been presented by
Elitzur and Vaidman  \cite{el-vaid,vaidman:94,bennett:95}. They have
shown that the presence of
an object can be detected without interacting with the object by making
use of a Mach-Zehnder interferometer. This interaction-free measurement
scheme has been optimized  and realized in an experiment performed by
the group of Zeilinger in Innsbruck \cite{zeilinger} \cite{zeilinger1}.

\section*{The Gedanken-experiment}
\noindent
In its simplest form, an interaction-free measurement can be made with
a Mach-Zehnder interferometer (cf. Figure 1). When this kind of
interferometer is empty, the amplitudes leading to detector $D_2$
interfere destructively and therefore only detector $D_1$ can fire.
If we insert into path $I$ an object which is assumed to be a perfect
absorber, this path is blocked and no interference can occur. Then
both detectors will fire with equal probability. Thus, if a single photon
is sent into the interferometer and a click is detected in $D_2$, one
knows with certainty
that an object is present in path $I$ without having interacted with
this
object. Of course it is also possible that the photon is absorbed by the
object or detected in $D_1$ but nevertheless in 25\% of all
trials we will succeed in  performing an interaction-free measurement.
With a more complicated setup
the percentage of successful trials can come arbitrarily close to
100\% \cite{zeilinger}.
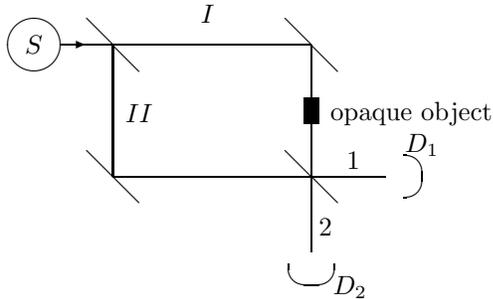
\begin{figure}
\begin{center}
\unitlength=0.70mm
\special{em:linewidth 0.4pt}
\linethickness{0.4pt}
\begin{picture}(83.67,61.00)
\put(70.67,33.00){\makebox(0,0)[cc]{$1$}}
\put(65.33,20.33){\makebox(0,0)[cc]{$2$}}
\put(10.00,55.00){\makebox(0,0)[cc]{$S$}}
\put(10.00,55.00){\circle{10.00}}
\put(80.17,30.00){\oval(7.00,8.00)[r]}
\put(83.67,36.00){\makebox(0,0)[cc]{$D_1$}}
\put(30.00,42.00){\makebox(0,0)[cc]{$II$}}
\put(62.67,13.33){\oval(8.67,8.00)[b]}
\put(70.00,9.00){\makebox(0,0)[cc]{$D_2$}}
\put(43.00,61.00){\makebox(0,0)[cc]{$I$}}
\put(15.00,55.00){\line(1,0){47.67}}
\put(62.67,55.00){\line(0,-1){39.33}}
\put(76.67,30.00){\line(-1,0){51.67}}
\put(25.00,30.00){\line(0,1){25.00}}
\put(20.00,60.00){\line(1,-1){10.00}}
\put(57.67,35.00){\line(1,-1){10.00}}
\put(57.67,60.00){\line(1,-1){10.00}}
\put(20.00,35.00){\line(1,-1){10.00}}
\put(61.33,40.00){\rule{2.67\unitlength}{5.00\unitlength}}
\put(66.33,42.00){\makebox(0,0)[lc]{opaque object}}
\put(17.00,55.00){\vector(1,0){2.99}}
\end{picture}
\end{center}
\caption{A Mach-Zehnder interferometer is shown. $S$ denotes the
source, $D_1$ and $D_2$ are detectors. If there is no absorber (opaque
object) present in path $I$ output 2 is dark and only detector $D_1$
fires. As soon as path $I$ is blocked both detectors can fire. In case
$D_2$ fires one knows that an absorber is present in path $I$ without
having interacted with it.}
\end{figure}

We now turn to our method of interaction-free preparation of a narrow atom
beam from an originally wide one.
Consider the Mach-Zehnder interferometer for neutrons shown in figure 2.
Let $w_n$ be the width of the beams inside the interferometer.
In path $I$ the neutron beam propagating along
the x-direction is crossed by a beam of $^{157}Gd$-atoms which is parallel to
the z-direction and has the width $w_{Gd}$ (cf. Figure 2).
We use $^{157}Gd$-atoms, because they are highly efficient neutron absorbers.
The $Gd$ beam is assumed to be
wider than the neutron beam $(w_{Gd} \gg w_n)$.
Without the atom beam all neutrons passing through the interferometer
will be detected in $D_1$.
As soon as we turn on the $Gd$-beam,
path $I$ of the neutron interferometer will be blocked once in a while
by an atom, which acts as a neutron absorber.
Then also detector $D_2$ can fire.
If it fires one knows that the $Gd$-atom was present within the region
of width $w_n$ defined by the neutron beam in path $I$.
Because path $I$ was blocked by the $Gd$-atom one also knows that the neutron
detected in $D_2$ took path $II$ and therefore never interacted with the
$Gd$-atom.
Interaction-free preparation of a $Gd$-beam of width $w_n$ from an originally
much wider beam is thus possible by installing a shutter for the $Gd$-beam
after the overlap with the neutron beam in path $I$.
This shutter opens only - with a suitable time delay - when a neutron
is detected in $D_2$, thereby permitting the selected $Gd$-atom to pass on.
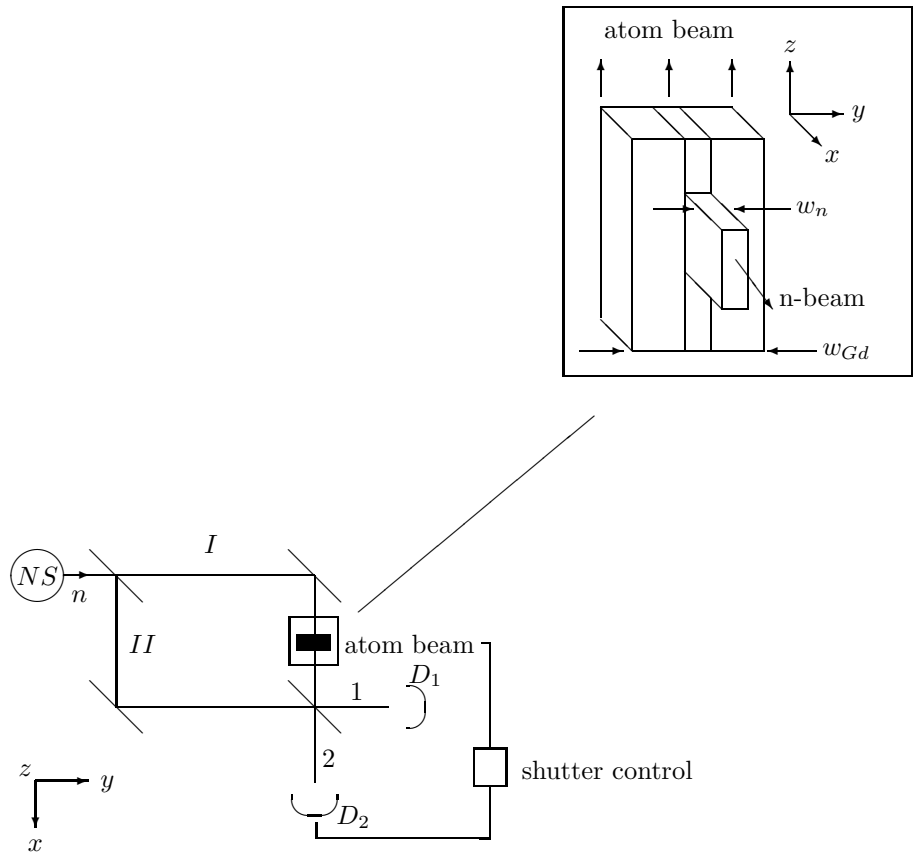
\begin{figure}
\begin{center}
\unitlength=0.70mm
\special{em:linewidth 0.4pt}
\linethickness{0.4pt}
\begin{picture}(176.00,163.00)
\put(10.00,55.00){\makebox(0,0)[cc]{$NS$}}
\put(10.00,55.00){\circle{10.00}}
\put(80.17,30.00){\oval(7.00,8.00)[r]}
\put(83.67,36.00){\makebox(0,0)[cc]{$D_1$}}
\put(30.00,42.00){\makebox(0,0)[cc]{$II$}}
\put(18.00,51.00){\makebox(0,0)[cc]{$n$}}
\put(70.67,33.00){\makebox(0,0)[cc]{$1$}}
\put(62.67,13.33){\oval(8.67,8.00)[b]}
\put(70.00,9.00){\makebox(0,0)[cc]{$D_2$}}
\put(65.33,20.33){\makebox(0,0)[cc]{$2$}}
\put(43.00,61.00){\makebox(0,0)[cc]{$I$}}
\put(15.00,55.00){\line(1,0){47.67}}
\put(62.67,55.00){\line(0,-1){39.33}}
\put(76.67,30.00){\line(-1,0){51.67}}
\put(25.00,30.00){\line(0,1){25.00}}
\put(20.00,60.00){\line(1,-1){10.00}}
\put(57.67,35.00){\line(1,-1){10.00}}
\put(57.67,60.00){\line(1,-1){10.00}}
\put(20.00,35.00){\line(1,-1){10.00}}
\put(68.33,42.00){\makebox(0,0)[lc]{atom beam}}
\put(58.00,38.00){\framebox(9.00,9.00)[cc]{}}
\put(17.00,55.00){\vector(1,0){2.99}}
\put(63.00,8.00){\line(0,-1){3.07}}
\put(63.07,5.08){\line(1,0){30.93}}
\put(96.00,5.08){\line(0,1){9.98}}
\put(96.00,21.94){\line(0,1){20.13}}
\put(93.00,15.00){\framebox(6.00,7.00)[cc]{}}
\put(102.00,18.00){\makebox(0,0)[lc]{shutter control}}
\put(59.33,40.67){\rule{6.33\unitlength}{3.00\unitlength}}
\put(9.67,16.00){\vector(0,-1){9.00}}
\put(9.67,16.00){\vector(1,0){10.00}}
\put(7.67,18.00){\makebox(0,0)[cc]{$z$}}
\put(23.33,15.33){\makebox(0,0)[cc]{$y$}}
\put(9.67,3.67){\makebox(0,0)[cc]{$x$}}
\put(117.00,104.00){\line(0,1){40.00}}
\put(117.00,144.00){\line(1,-1){6.01}}
\put(123.00,137.67){\line(0,-1){40.00}}
\put(123.00,97.67){\line(-1,1){6.01}}
\put(123.00,97.67){\line(1,0){25.00}}
\put(148.00,97.67){\line(0,1){40.00}}
\put(148.00,137.92){\line(-1,0){25.00}}
\put(117.00,144.00){\line(1,0){25.00}}
\put(142.00,144.00){\line(1,-1){6.01}}
\put(145.00,120.67){\line(0,-1){15.00}}
\put(140.00,105.67){\line(0,1){15.00}}
\put(145.00,120.67){\line(-1,1){7.00}}
\put(140.00,120.67){\line(-1,1){7.00}}
\put(140.00,105.67){\line(-1,1){7.00}}
\put(133.00,112.67){\line(0,1){15.00}}
\put(127.00,144.00){\line(1,-1){6.01}}
\put(132.00,144.00){\line(1,-1){6.01}}
\put(133.00,137.67){\line(0,-1){40.00}}
\put(138.00,137.67){\line(0,-1){10.00}}
\put(138.00,107.67){\line(0,-1){10.00}}
\put(153.00,124.67){\vector(-1,0){10.67}}
\put(127.00,124.67){\vector(1,0){7.67}}
\put(157.34,124.67){\makebox(0,0)[cc]{$w_n$}}
\put(113.00,97.67){\vector(1,0){8.33}}
\put(158.00,97.67){\vector(-1,0){9.33}}
\put(163.67,97.67){\makebox(0,0)[cc]{$w_{Gd}$}}
\put(153.00,142.67){\vector(0,1){10.00}}
\put(153.00,142.67){\vector(1,0){10.00}}
\put(153.00,142.67){\vector(1,-1){6.01}}
\put(153.00,155.67){\makebox(0,0)[cc]{$z$}}
\put(166.00,142.67){\makebox(0,0)[cc]{$y$}}
\put(161.00,134.67){\makebox(0,0)[cc]{$x$}}
\put(142.67,115.00){\vector(3,-4){7.00}}
\put(133.00,127.67){\line(1,0){5.00}}
\put(140.00,120.67){\line(1,0){5.00}}
\put(140.00,105.67){\line(1,0){5.00}}
\put(137.98,137.22){\line(0,1){0.67}}
\put(132.99,137.39){\line(0,1){0.50}}
\put(122.99,137.14){\line(0,1){0.75}}
\put(148.02,137.24){\line(0,1){0.66}}
\put(150.90,108.02){\makebox(0,0)[lc]{n-beam}}
\put(96.04,42.16){\line(-1,0){1.60}}
\put(130.00,159.00){\makebox(0,0)[cc]{atom beam}}
\put(110.00,93.00){\framebox(66.00,70.00)[cc]{}}
\put(117.00,146.00){\vector(0,1){7.00}}
\put(142.00,146.00){\vector(0,1){7.00}}
\put(130.00,146.00){\vector(0,1){7.00}}
\put(71.00,48.00){\line(6,5){44.83}}
\put(94.00,5.00){\line(1,0){2.02}}
\end{picture}
\end{center}
\caption{Neutrons from the source $NS$ are incident on a Mach-Zehnder
neutron interferometer and can finally be registered by detectors
$D_1$ and $D_2$. In path $I$ the neutron
beam propagating along the x-direction is crossed by a $^{157}Gd$-atom beam
which is parallel to the z-direction. It is assumed that the atom beam is
wider than the neutron beam as shown in the detail clipping.}

\end{figure}

\section*{Formal description}

We now turn to a more detailed discussion.
For the sake of simplicity the neutron beam is assumed to be of rectangular
cross section with constant transverse probability density. This comes close
to real experimental conditions.
An analogous assumption is made for the
atom beam. For the following it is sufficient to consider only one transverse
direction of the beams. Similarly, their longitudinal description can be
ignored. Because $w_{Gd} \gg w_n$ is assumed,
we represent the transverse probability density of the atom beam in real
space as a superposition of a rectangular wavepacket
$|a \rangle_R$ of width $w_n$, which exactly crosses the neutron beam
(cf. figure 2), and of another wavepacket $|a \rangle_0$
which represents the rest of the beam:
\[
|a \rangle = |a \rangle_R + |a \rangle_0
\]
When the neutron and atom wavepackets overlap the following processes
can happen:
\newcounter{zahl}
\begin{list}{(\roman{zahl})}{\usecounter{zahl}}
\item
Neutron and atom don\rq t interact.
\item
The neutron is scattered by the atom.
\item
The neutron is absorbed by the atom.
\end{list}
Corresponding to these possibilities the combined state of the atom
and of the neutron in
path $I$ after the overlap is given by
\begin{eqnarray}
|n \rangle_I |a \rangle &=& |n \rangle_I \left(|a \rangle_0
+ c |a \rangle_R\right)
\nonumber   \\
&& + \sum_l s_l |n \rangle_{I,l} |a \rangle_{R,l}  \nonumber \\
&& + z |n \rangle_I |a \rangle_R .
\label{moeglich}
\end{eqnarray}
Here $s_l$ are the probability amplitudes for scattering, where
$l$ labels the exchange of momentum and kinetic energy
between neutron and atom, and therefore also appears in the
resulting state vectors.
$s_0$ is the amplitude for forward scattering,
which neither changes the state
of the neutron nor that of the atom, but adds a phase factor.
The absorption amplitude is given by $z$.
The amplitude $c$ in the first term on the right hand side
expresses the probability that the $Gd$-atom crosses through the neutron
beam without scattering or absorption, and is given by
\[
c = \sqrt{1 - \sum_l |s_l|^2 - |z|^2}.
\]
Note, that for the interaction of slow neutrons with $^{157}Gd$, scattering
is four orders of magnitude less likely than absorption, because the
cross section for absorption is $2.5\times 10^5$ barns ($10^{-24} cm^2$)
whereas that for scattering is of the order of 10 barns. (No exact value
is known for $^{157}Gd$. The value for natural gadolinium, which contains
15.65\% of $^{157}Gd$, is 7 barns.) Consequently we have
$\sum_l |s_l|^2 \ll |z|^2$.

Now the probability amplitude for detection of a neutron in $D_2$ can be
calculated. The neutron can reach detector $D_2$ by the following routes:
\newcounter{number}
\begin{list}{(\roman{number})}{\usecounter{number}}
\item
It passes through the interferometer along path $II$.
\item
It passes through the interferometer along path $I$ and is neither
scattered (except forward scattering) nor absorbed.
\end{list}
Thus we get for the combined state of the neutron just before detector $D_2$,
and of the atom:
\begin{eqnarray}
|n \rangle_{D_2} |a \rangle  & = &  |n \rangle_{II,D_2}
|a \rangle  \nonumber \\
&& + |n \rangle_{I,D_2} |a \rangle_0 +
     c |n \rangle_{I,D_2} |a \rangle_R +
     s_0 |n \rangle_{I,D_2} |a \rangle_R.
      \label{result}
\end{eqnarray}
The two states contributing to the output of the Mach-Zehnder interferometer
towards detector $D_2$ are functions of the input state $|n \rangle_0$.
Neglecting the directions of the beams, these states are given by
\begin{eqnarray}
|n \rangle_{I,D_2}  & = & \frac{i}{2} |n \rangle_0  \nonumber \\
|n \rangle_{II,D_2} & = & -\frac{i}{2} |n \rangle_0,  \label{ifmout}
\end{eqnarray}
such that eq.(\ref{result}) can be rewritten as
\begin{equation}
|n\rangle_{D_2}|a\rangle = \frac{i}{2}(c+s_0-1)|n\rangle_0|a\rangle_R.
              \label{exact}
\end{equation}
With realistic dimensions of the beam width $w_n$, from a few $\mu m$
upwards, and with the usual very sparse beams, most of the time
the neutrons and the $Gd$-atoms will not interact $(c \approx 1)$. But in
the rare cases when an interaction occurs it is predominately absorption
because of $|z|^2 \gg \sum_l |s_l|^2$, and $|z|^2 \gg |s_0|$. Therefore
eq.(\ref{exact}) reduces to
\begin{equation}
|n \rangle_{D_2} |a \rangle \approx
 - \frac{i}{4} |z|^2 |n \rangle_0 |a \rangle_R.
\label{ergebnis}
\end{equation}
Equation (\ref{ergebnis}) expresses the fact that by detecting a
neutron in $D_2$
one has reduced the original state of the atom
$|a \rangle = |a \rangle_R + |a \rangle_0$ to $|a \rangle_R$.
The atom is thus indeed confined to a wavepacket,
which has the width of the neutron beam.
This corresponds to a gain of knowledge about the position of the atom.
Because the neutron by which this gain has been reached almost always
took path $II$ and therefore cannot have interacted with the atom,
this is a method of preparing the state $|a \rangle_R$ without any physical
interaction.

\section*{Discussion}
\noindent
Our analysis has shown that information about the presence of
an atom can apparently be obtained without interaction and can be used
for  preparing an atomic beam  as narrow as the "probing" neutron beam.
In view of the original treatment of interaction free measurement by Dicke
\cite{dicke}, which  concluded that the absence of
interaction was only an illusion,
it is worth while to investigate in what sense there was no interaction
between the neutrons and those atoms, which ultimately compose the
narrow beam.

Dicke considered the Gaussian wavepacket of an atom traversed by
a beam of light much narrower than the wavepacket.
Photons scattered at the atom are detected, whereby one learns
that the atom was within the beam of light. If no scattered photons
are detected with a properly adjusted intensity of the light, one learns
that the atom is outside the beam of light. This information seems to have
been gained without interaction. It yields a new
wavefunction showing the atom localized somewhere in a ring around
the beam of light. This is a narrower structure than the atom's
original wavepacket, with a corresponding increase of the kinetic energy.
Where could this additional energy have come from? By means of
perturbation theory Dicke ascribes this to the
absorption and reemission of a photon by the atom. The reemitted photon
is not detected as scattered, because it is within the momentum
uncertainty of the focussed beam of light. Thus the measurement result
"atom is outside the beam of light" is only apparently obtained without
interaction.

There are two essential differences between the measurement scheme
discussed by Dicke and the preparation method presented here. One is
that in our setup information about the system is gained
by means of interference. The other is that in addition to scattering we
also consider absorption.
Nevertheless all effects discussed by Dicke are relevant in order
to describe what happens between the atoms and neutron beam I.

If there were only neutron beam I, we could observe the absorption of a
neutron by an atom by detecting the high energy photon emitted by the atom
in the transformation processes of the nucleus, or we could detect the
scattered neutron. Both processes would correspond to the detection of a
photon scattered by the atom in the case discussed by Dicke. If the neutron
is not absorbed by an atom, it did not "see" the atom, or it was
scattered in forward direction accounted for by the amplitude $s_0$
in eq.(1). Forward scattering is the interaction analogous to
scattering within the momentum uncertainty of the beam of light in Dicke's
case.

Thus, without an interference loop for the neutron, information about the
localization of the atom would be obtained in a similar way as in
Dicke's case:
Detection of a high energy photon, or of the scattered neutron, would
indicate that the atom was within the width of the neutron beam.
If neither effect is present, we obtain a new wavefunction for the atom,
which has a smaller amplitude in the region crossed by the neutron beam.

{\it With} the interference loop, however,
the absence of a high energy photon or a scattered neutron may result in
{\it two} different informations, as either $D_1$
or $D_2$ may fire. The firing of $D_1$ tells us little about
the new wavefunction for the atom. But when $D_2$ fires we can think of two
different causes:

\noindent
1.) It is either due
to the phase shift the neutron acquired in forward scattering at the atom,
and hence due to an interaction in the region of the neutron beam.
This localizes the atom within the width of neutron beam I. Naturally,
it cannot be said whether the interaction has actually taken place,
as path II is also open to the neutron.

\noindent
2.) Or, it is due to the atom acting as an absorber and blocking path I.
But rather than being really absorbed, the neutron took the other path
available in the interferometer. This, too, localizes the
atom within the width of neutron beam I, which could be interpreted as
gaining information by "frustrated absorption".
For the parameters in our example
this is by far the main reason for a firing of $D_2$.

In this context two facts are important.
First, it should be noted that in Dicke's case
non-detection of a scattered photon localizes the atom {\it outside} the
beam  of light, whereas in our case the analogous
processes localize the atom {\it within} the neutron beam.
And second, the possible absorption of the neutron does remain an
{\it unused} possibility, because otherwise one would have detected a high
energy photon. One can only conclude that these neutrons have come along
path II, and hence have not interacted with the atoms.
In Dicke's case nothing analogous can be found.

The narrow beam of selected atoms must fulfill the Heisenberg
uncertainty relations, which require the transverse kinetic energy to have
increased. For most of the atoms this cannot have happened as they are
selected  by "frustrated absorption", where really no interaction seems
to occur. Only the contribution from forward scattering leaves room for the
exchange of energy from the neutron to the atom. This can indeed account for
the necessary change of energy, because according to the optical theorem, the
total reaction cross section, which in our case includes scattering and
absorption, is proportional to the amplitude for forward scattering.


\begin{thebibliography}{99}

\bibitem{Gabor}
D. Gabor ,{\sl Progress in Optics} {\bf 1}, 111 - 153 (1961).

\bibitem{el-vaid}
A. C. Elitzur and L. Vaidman, {\sl Foundations of Physics} {\bf 23}, 987
(1993).

\bibitem{vaidman:94}
L. Vaidman, {\sl Quantum optics} {\bf 6}, 119 (1994).

\bibitem{bennett:95}
Ch. Bennett,
{\sl Nature} {\bf 371}, 479 (1994).

\bibitem{zeilinger} P.Kwiat,H.Weinfurter,T.Herzog,A.Zeilinger,M.Kasevich,
{\sl Annals of the New York Academy of Sciences} {\bf 755}, 383-394 (1995).

\bibitem{zeilinger1} P.Kwiat,H.Weinfurter,T.Herzog,A.Zeilinger,M.Kasevich,
``Interaction-free Measurement'', {\sl Phys.Rev.Lett.}, in print.

\bibitem{dicke}
R.H. Dicke, {\sl Am.J.Phys.} {\bf 49}(10), 925-930 (1981).

\end{thebibliography}
\end{document}